\documentclass[journal,dvipsnames]{IEEEtran}
\usepackage{amsmath,amsfonts,amssymb,mathtools}
\usepackage{etex}
\usepackage{cite}
\ifCLASSINFOpdf
\usepackage{graphicx}
\else
\fi

\usepackage{empheq}
\usepackage{textcomp}
\usepackage{tikz}
\usepackage{pgfplots}
\pgfplotsset{compat=1.8}
\usetikzlibrary{arrows,decorations.markings}

\usepackage{multicol}

\usepackage{verbatim}
\usetikzlibrary{matrix,calc,shapes}
\usetikzlibrary{3d}
\usetikzlibrary{positioning}
\usetikzlibrary{shapes,shadows,calc}
\usepgflibrary{arrows}

\definecolor{mycolor1}{HTML}{CA0020}
\definecolor{mycolor2}{HTML}{F4A582}
\definecolor{mycolor3}{HTML}{F7F7F7}
\definecolor{mycolor4}{HTML}{92C5DE}
\definecolor{mycolor5}{HTML}{0571B0}

\tikzset{
  treenode/.style = {shape=rectangle, rounded corners,
                     draw, anchor=center,
                     text width=5em, align=center,
                     top color=white, bottom color=blue!20,
                     inner sep=1ex},
  decision/.style = {treenode, diamond, inner sep=0pt},
  root/.style     = {treenode, font=\Large, bottom color=red!30},
  env/.style      = {treenode, font=\ttfamily\normalsize},
  finish/.style   = {root, bottom color=green!40},
  dummy/.style    = {circle,draw}
}

\tikzset{
  sshadow/.style={opacity=.25, shadow xshift=0.05, shadow yshift=-0.06},
}

\def\aboxl[#1,#2,#3,#4,#5]#6{%
  \node[draw, cylinder, alias=cyl, shape border rotate=0, aspect=#3, %
  minimum height=#1, minimum width=#2, outer sep=-0.5\pgflinewidth, %
  color=mycolor3, top color=white, bottom color=gray, middle
  color=mycolor3] (#4) at #5 {};%
  \node at #5 {#6};%
  \fill [blue]
  } 
\def\aboxr[#1,#2,#3,#4,#5]#6{%
  \node[draw, cylinder, alias=cyl, shape border rotate=0, aspect=#3, %
  minimum height=#1, minimum width=#2, outer sep=-0.5\pgflinewidth, %
  color=mycolor3, top color=white, bottom color=gray, middle
  color=mycolor3] (#4) at #5 {};%
  \node at #5 {#6};%
  \fill [blue]
  }
  \def\aboxm[#1,#2,#3,#4,#5]#6{%
  \node[draw, cylinder, alias=cyl, shape border rotate=0, aspect=#3, %
  minimum height=#1, minimum width=#2, outer sep=-0.5\pgflinewidth, %
  color=mycolor2, top color=white, bottom color=gray, middle
  color=mycolor1] (#4) at #5 {};%
  \node at #5 {#6};%
  \fill [red] 
}
\def\aboxrv[#1,#2,#3,#4,#5]#6{%
  \node[draw, cylinder, alias=cyl, shape border rotate=90, aspect=#3, %
  minimum height=#1, minimum width=#2, outer sep=-0.5\pgflinewidth, %
  color=mycolor3, left color=mycolor3, right color=mycolor3, middle
  color=mycolor3] (#4) at #5 {};%
  \node at #5 {#6};%
  \fill [blue]
  }
   
\def\aboxmv[#1,#2,#3,#4,#5]#6{%
  \node[draw, cylinder, alias=cyl, shape border rotate=90, aspect=#3, %
  minimum height=#1, minimum width=#2, outer sep=-0.5\pgflinewidth, %
  color=mycolor2, left color=white, right color=white, middle
  color=mycolor1] (#4) at #5 {};%
  \node at #5 {#6};%
  \fill [red] 
}
\usepackage{pgfplotstable}
\usepackage{booktabs}

\newcommand{\fref}[1]{Fig.~\ref{#1}}

\def\minim{\mathop{\hbox{\rm minimize}}} \def\subject{\text{\rm subject to}} \def\minimize#1{{\displaystyle\minim_{#1}}}

\usepackage{hyperref}

\newcommand{\fsum}[3]{\sum_{#1}^{#2}{#3}}
\newcommand{\pder}[2]{\frac{\partial #1}{\partial #2}}

\def\u{{\mathbf u}}

\def\x{{\mathbf x}}

\def\p{{\mathbf{g}}}

\def\0{{\mathbf 0}}
\def\blambda{{\pmb{\lambda}}}

\def\F{f}

\def\myobj{J}
\def\optwidth{0.4}

\def\minim{\mathop{\hbox{\rm minimize}}}
\def\minimize#1{{\displaystyle\minim_{#1}}}
\def\subject{\text{\rm subject to}}

    \def\probNL#1#2#3#4#5#6#7#8{
   {\begin{tabular*}{\optwidth\textwidth}
    {@{}l@{\extracolsep{4pt}}l@{\extracolsep{4pt}}l@{\extracolsep{\fill}}l@{}}
         $\minimize{#2}$ & $#3$                 \\[4pt]
         $\subject$      & $#4$ \hfill{~} (#1)  \\[4pt]
                         & $#5$                 \\[4pt]
                         & $#6$  $#8$ \; $#7$ \\[4pt]
    \end{tabular*}}}

\def\probTL#1#2#3#4#5#6#7#8#9{
   {\begin{tabular*}{\optwidth\textwidth}
    {@{}l@{\extracolsep{4pt}}l@{\extracolsep{4pt}}l@{\extracolsep{\fill}}l@{}}
         $\minimize{#2}$ & $#3$                         \\[4pt]
         $\subject$      & $#4$                         \\[4pt]
                         & $#5$ \hfill{~} (#1)          \\[4pt]
                         & $#6$                         \\[4pt]
                         & $#7$  $#9, \; \; #8$    \\[4pt]
\end{tabular*}}}

\makeatletter
\newcommand*{\transpose}{%
	  {\mathpalette\@transpose{}}%
}
\newcommand*{\@transpose}[2]{%
	        \raisebox{\depth}{$\m@th#1\intercal$}%
}
\makeatother

\ifCLASSINFOpdf
\else
\fi

\usepackage{algpseudocode}
\usepackage{algorithm}

\algnewcommand{\algorithmicand}{\textbf{ and }}
\algnewcommand{\algorithmicor}{\textbf{ or }}
\algnewcommand{\OR}{\algorithmicor}
\algnewcommand{\AND}{\algorithmicand}
\algnewcommand{\TRUE}{\textbf{ true }}
\algnewcommand{\FALSE}{\textbf{ false }}
\algnewcommand{\NOT}{\textbf{ not }}
\algnewcommand{\BREAK}{\textbf{ break }}
\algrenewcommand\alglinenumber[1]{{\sffamily\footnotesize#1}}

\begin{document}
%
\title{Optimizing gas networks using adjoint gradients}


\author{Conor~O'Malley\IEEEauthorrefmark{2},
        Drosos~Kourounis\IEEEauthorrefmark{1},
        Gabriela~Hug\IEEEauthorrefmark{2},
        and~Olaf~Schenk\IEEEauthorrefmark{1},
\thanks{\IEEEauthorrefmark{1}Power Systems Laboratory, ETH Zurich, Switzerland.}
\thanks{\IEEEauthorrefmark{3}Advanced Computing Laboratory, Institute of Computational Science, USI Lugano, Switzerland.}
\thanks{This work is funded by Commission for Technology and Innovation Switzerland: Project No. 18801.1 PFIW-IW}}

\maketitle

\begin{abstract}
	An increasing amount of gas-fired power plants are currently being installed in modern power grids worldwide. This is due to their low cost and the inherent flexibility offered to the electrical network, particularly in the face of increasing renewable generation. 
	However, the integration and operation of gas generators poses additional challenges to gas network operators, mainly because they can induce rapid changes in the demand. 
	This paper presents an efficient minimization scheme of gas compression costs under dynamic conditions where deliveries to customers are described by time-dependent mass flows. The optimization scheme is comprised of a set of transient nonlinear partial differential equations that model the isothermal gas flow in pipes, an adjoint problem for efficient calculation of the objective gradients and constraint Jacobians, and state-of-the-art optimal control methods for solving nonlinear programs. As the evaluation of constraint Jacobians can become computationally costly as the number of constraints increases, efficient constraint lumping schemes are proposed and investigated with respect to accuracy and performance.
	The resulting optimal control problems are solved using both interior-point and sequential quadratic programming methods. The proposed optimization framework is validated through several benchmark cases of increasing complexity. 

\end{abstract}



%
\IEEEpeerreviewmaketitle

\section{Introduction}
In recent years, there has been a global increase in the production of natural gas due to the advent of hydraulic fracturing to recover natural gas from shale rock formations. The abundance of natural gas combined with its competitively low cost, and its lower carbon intensity compared to coal has led to an increase in the amount of gas-fired generation units in the electrical system. At the same time there is an increasing penetration of renewable energy sources (RES) in the electrical system. Any fluctuation caused by these RES needs to be balanced by other resources. Certain types of gas-fired generation are well suited for these balancing actions, however, the rapid change of gas demand can create challenging operating conditions for the gas network operator.

Historically, the flows through the network were time invariant allowing the assumption of steady state flow. However, the current paradigm shift towards more variable gas-fired generator operation falsifies this assumption and requires transient state modelling of the gas network. Modelling the transient state of the network is computationally expensive. However, it can provide much more detailed insights into the network state at any time instance. Another benefit of modelling the transient state of the network is that the effect of the networks linepack are naturally incorporated into the model~\cite{bal2016}.


The transients in the network are the outcome of the inflows and outflows in the network but also the setting of certain active network elements such as valves and compressors. The set points for these active elements can be chosen based on operator experience and rule-based control schemes or preferably by solving an optimal control problem of either the steady state problem \cite{Misra:2015} or the transient problem\cite{mak2016efficient} ensuring reliable and cost effective network operation. 
This is especially important given the increasing variability in gas demand due to uncertainty in the electrical system which requires the selection of set points that are robust against possible future outcomes. There are many techniques~\cite{Cosmin} for optimization under uncertainty, however, a limiting factor in applying these techniques for intercoupled energy systems, is the computational intensity of solving the optimal control problem for a transient gas network model. 
While much research has been done on the optimized operation of electrical networks\cite{kourounis2018towards}, optimization of gas networks and combined gas-electric networks has only recently gained significant attention\cite{zlotnik2017}\cite{wang2017convex}.

This paper introduces an efficient treatment of the optimal gas flow (OGF) problem where the minimization of gas compression costs are subject to dynamic equality and inequality constraints. The equality constraints are the isothermal transient partial differential equations (Euler's equations), introduced in Sect.~\ref{sec:GasNetwork}, that model the gas flows in pipes. The inequality constraints guarantee reliable and secure network operation by limiting the pressure at every node to be between operational limits.
The required discretization of the aforementioned PDEs both in space and time is described in Sect.~\ref{sec:Discretization}. Furthermore, a Newton-continuation scheme is introduced for robust convergence at each timestep under discontinuous compressor ratios. 
A detailed description of the OGF problem for minimizing gas compression costs subject to the aforementioned constraints
is provided in Sect.~\ref{sec:OGF}. This is followed by the introduction of a discrete adjoint problem formulation for the efficient calculation of the objective gradients and constraint Jacobians and efficient constraint-lumping schemes, introduced in~\cite{Kourounis2014,Kourounis2015}, for the acceleration of the solution of the OGF problem.
State-of-the-art interior-point and sequential quadratic programming methods are used to guide the minimization of the objective using the gradients provided from the adjoint problem. Through several benchmark cases of increasing complexity we investigate the efficiency and robustness of the proposed constraint lumping techniques in Sect. \ref{sec:Results}  and we conclude in Sect.~\ref{sec:conclusion}.

\section{Gas Network Modelling}
\label{sec:GasNetwork}

A gas network consists of supply nodes and demand nodes, interconnected primarily with pipelines. The gas flow in a pipeline is driven by the pressure difference across the pipeline. 
When using steady state assumptions for the gas network, the resulting model simplifies to a nonlinear algebraic set of equations. This computationally simpler steady-state analysis is frequently used for long-term problems such as gas network planning, and have been used in many works on gas-electric interactions\cite{Misra:2015,Shabanpour:2016}. However, transients in the gas network can be on the order of hours and the network may never reach steady state. Therefore, it is important to properly account for these transients in daily operational planning by modeling the system with Euler's equations for compressible fluids.

\subsection{Isothermal gas flow in pipelines}
Under the assumptions of a one-dimensional pipe domain, single phase flow, constant temperature, and steady state friction factor $f_r$ calculated by the Chen's formula~\cite{Chen}, the Euler equations are simplified to
\begin{subequations}
	\label{eq:Euler}
	\begin{align}
		\label{eq:MassBalanceEquation}
		&\frac{\partial \rho}{ \partial t} + \frac{\partial (\rho u)}{ \partial x} = 0, \\
		\label{eq:MomentumBalanceEquation}
		&\frac{\partial (\rho  u)}{ \partial t} +  \frac{\partial (\rho  u^2 + p(\rho))}{ \partial x} = - f_r \frac{\rho u |\rho u|}{2D \rho},
	\end{align}
\end{subequations}
where $D$ is the pipe diameter, $\rho(x, t)$ is the gas density and $u(x, t)$ is the velocity of gas. In high-pressure gas systems  where $u \ll c$, the term $\partial (\rho u^2)$ can be neglected\cite{Osiadacz:1996}. The equation of state $p(\rho)$ describing the dependence of pressure $p$ on the density $\rho$ is, in the context of isothermal gas flow, usually assumed as
\begin{align}
	\label{eq:EquationOfState}
	p(\rho) = \frac{z \mathcal{R} T}{M_g}\rho = c^2 \rho,
\end{align}
where $z$ is the natural gas compressibility factor, $\mathcal R$ the universal 
gas constant, $M_g$ the molecular weight of the gas, and $T$ the temperature of the gas. The parameter $c$ can be viewed as the speed of sound in the gas. Using the fact that the mass flow $m$ is
\begin{align}
	\label{eq:massflow}
	m=\rho u \frac{\pi D^2}{4}
\end{align}
and incorporating this into \eqref{eq:Euler} and \eqref{eq:EquationOfState}  leads to the following set of equations
\begin{subequations}
	\label{eq:IsothermalGasFlowEquations}
	\begin{align}
		\label{eq:IsothermalGasFlowEquations1}
		&\frac{\partial p}{ \partial t} =-\frac{c^2}{A}\frac{\partial  {m}}{ \partial x}, \\ 
		\label{eq:IsothermalGasFlowEquations2}&\frac{\partial {m}}{ \partial t} +A\frac{\partial p}{ \partial x} =\frac{-f_r c^2  {m} | {m}|}{2DAp}.
	\end{align}
\end{subequations}

\subsection{Compressor stations}
In order to transport gas over large distances compressor stations are required to overcome the pressure loss that occurs due to the friction in the pipes. Here a multiplicative compression model is used which is described with the following equations\cite{zlotnik2017}. The change in pressure from the inlet to the outlet of a compressor is given by:
\begin{align}
	\kappa=\frac{P_{out}}{P_{in}},
	\label{eq:CompressorRatioRelationship}
\end{align}
where $\kappa$ is the compression ratio, $P_{out}$ and $P_{in}$ are the outlet and inlet pressures, respectively.

A compressor must be driven by a motor that can be powered by siphoning off an amount of the gas passing through the compressor. The amount of gas required for a given compression ratio is
\begin{equation}
\label{eq:CompressorConsumption}
m_{con}=Km_{out}\big(\kappa^{\gamma}-1\big),
\end{equation}
where $m_{con}$ is the mass flow consumed by the compressor station, $m_{out}$ is the mass flow out of the compressor station, $K$ and $\gamma$ are constants based on the characteristics of the compressor station and the gas.

\section{Discretization}
\label{sec:Discretization}
In order to solve the system of nonlinear equations \eqref{eq:IsothermalGasFlowEquations}, the partial differential equations need to be discretized. Several high resolution schemes have been suggested for solving Euler equations, see, e.g.,~\cite{Leveque1992numerical,Leveque2002finite}.
Here, we will opt for the cell-centered 
finite volume method presented in \cite{Helgaker:2014}, which is formulated through 
differentiable fluxes in contrast to  high resolution methods involving nondifferentiable flux limiters. Furthermore, it converges quadratically, 
and it allows straightforward integration of boundary conditions.

\subsection{Implicit Cell Centered Method}
Cell-centered finite volume methods are directly applicable on structured meshes
topologically equivalent to a uniform Cartesian grid and they can be trivially 
applied on a one-dimensional grid such as the gas pipe network. The partial space
derivatives of the isothermal gas flow equations 
\eqref{eq:IsothermalGasFlowEquations} are approximated by divided differences, 
and the resulting nonlinear system of algebraic equations needs to be solved at each time step.
\subsubsection{Gas Network Mesh}
\begin{figure}[!t]
	\centering
	\begin{tikzpicture}[
	>=stealth,
	iron/.style={shade, ball color=mycolor4},
	electron/.style={shade, ball color=black},
	oxygen/.style={shade, ball color=mycolor3},
	droplet/.style={ball color=mycolor3, opacity=0.4},
	]
	\small
	\aboxl[60,25,2.5,a1,(-1,0)] {\textcolor{black}{$I-1$}}; 
	\aboxm[30,25,2.5,a2,(0.1,0)] {};
	\aboxr[60,25,2.5,a2,(1.2,0)] {\textcolor{black}{$I$}};
	\aboxm[30,25,2.5,a2,(2.3,0)] {};
	\aboxr[60,25,2.5,a2,(3.4,0)] {\textcolor{black}{$I+1$}};
	\node at (0.2,-.8) {$x_i$};
	\node at (2.5,-.8) {$x_{i+1}$};
	
	\node  at (0.2,0.65) {$m_{i}$};
	\node  at (0.2,0.9) {$p_{i}$};
	
	\node  at (2.5,0.65) {$m_{i+1}$};
	\node  at (2.5,0.9) {$p_{i+1}$};
	
	\end{tikzpicture}
	\caption{Cell-centered discretization scheme: control volumes in grey and nodes in red}
	\label{fig:FDgrid}
\end{figure}
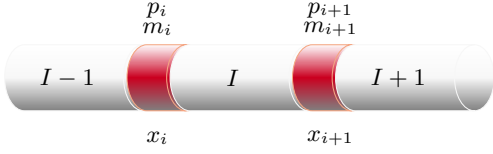

The physical gas network is described by a connected graph
consisting of a set of nodes $\mathcal{N}_0$ and edges $\mathcal{E}_0$ representing
the pipelines. Each pipeline is further subdivided into a set
of control volumes indicated by $\mathcal{E}$. Additional nodes 
are introduced by the subdivisions and  $\mathcal{N}$ indicates the 
set of these additional nodes together with the original nodes.
At each node $j \in \mathcal{N}$, both the pressure $p_j(t)$ and 
mass flux $m_j(t)$ variables are specified as shown in \fref{fig:FDgrid}.
Their values are computed using the cell-centered method to 
enforce equations \eqref{eq:IsothermalGasFlowEquations} at each of the control volumes.


In most implementations, the nodal value of the approximate solution at the $i$th node
$u_i(t) \approx u(x_i, t)$ is a pointwise approximation of the true
solution of the underlying PDE.  For gas flow equations, however, Helgaker et al.~\cite{Helgaker:2014}
uses nodal values to approximate average pressure and mass flux variables within each 
pipe subdivision of length $\delta x$. More precisely, the average value of these variables at the $n$th timestep and at the center $x_I$ of the $I$th control volume shown in \fref{fig:FDgrid}, is represented by $u(x, t^n)$ and approximated as

\begin{align}
	u_I^n = \frac{1}{2} (u^{n}_{i}+u^{n}_{i+1})
	+ \mathcal{O}(\delta x^2)
\end{align}
and the first order spatial derivative at $x_I$, is computed from
\begin{equation}
\frac{\partial u_I^n}{\partial x} = \frac{u^{n}_{i+1}-u^{n}_{i}}{\delta x}
+ \mathcal{O}(\delta x^2)
\end{equation}
The temporal derivative of $u$ at $x_I$ can be approximated by the backward Euler formula,
\begin{equation}
\frac{\partial u_I^n}{\partial t} 
= \frac{u_I^{n} - u_I^{n-1}}{\delta t} 
+ \mathcal{O}(\delta t). 
\end{equation}

Application of these discrete formulas to the Euler equations \eqref{eq:IsothermalGasFlowEquations} results in the discretized Euler equations for every $I \in \mathcal{E}$, given by 
\begin{subequations}
	\label{eq:GF}
	\begin{align}
		\label{eq:DiscreteMassBalance} 
		&\frac{1}{\delta t}\left ( p^n_{I}-p^{n-1}_{I}\right )+ 
		\frac{c^2}{A \delta x}\left ( m^n_{i+1}-m^n_{i,n} \right )=0,
		\\
		\label{eq:DiscreteMomentumBalance}
		&\frac{1}{\delta t}\left ( m^n_{I}-m^{n-1}_{I} \right) +\frac{A}{ \delta x}
		\left ( p^{n}_{i+1}-p^{n}_{i} \right )=\frac{-f_{r}c^2}{2DA}\cdot\frac{|m^n_{I}|m^n_{I}}{p^n_{I}}.
	\end{align}
\end{subequations}

This cell centered method is second order accurate in space, and since we adopt the backward Euler formula for the time discretization, it is first order accurate in time, leading to a nonlinear system of equations to be solved at each timestep.
For a pipeline consisting of $N_p$ grid points and $N_p-1$ control volumes, the cell centered method will lead to a nonlinear algebraic system of $2 N_p-2$ equations with $2 N_p$ unknowns. Therefore two unknowns must be included as boundary conditions. This is done by specifying one slack node with fixed pressure and the mass flux (i.e. supply or load) of the pipeline.
\subsubsection{Compressors}
\begin{figure}[t]
	\centering
	\begin{tikzpicture}[
	>=stealth,
	iron/.style={shade, ball color=red},
	electron/.style={shade, ball color=black},
	oxygen/.style={shade, ball color=blue},
	droplet/.style={ball color=blue!20, opacity=0.4},
	]
	
	\tikzset{xzplane/.style={canvas is xz plane at y=#1, thick,ball color=mycolor4,draw=mycolor4}};
	\tikzset{yzplane/.style={canvas is yz plane at x=#1,thick, ball color=mycolor4}};
	\tikzset{xyplane/.style={canvas is xy plane at z=#1,thick,shading=ball,ball color=mycolor4,draw=mycolor4}};
	
	\small
	
	\aboxr[60,25,2.5,a2,(2.0,0)] {\textcolor{black}{$I$}};
	\aboxm[30,25,2.5,a2,(3.0,0)] {};
	
	\begin{scope}[shift={(3.5 cm,-0.2 cm)}]
	\draw[xzplane=0.85] (0.0,0.0)--(1,0.15)--(1,0.85)--(0.0,0.85) --cycle;
	\draw[yzplane=1.] (0.15,0.15)--(0.85,0.15)--(0.85,0.85)--(0.15,0.85) --cycle;
	\draw[xyplane=0.85] (0.,0)--(1,0.15)--(1,0.85)--(0.,0.85) --cycle;
	\end{scope}
	\aboxm[25,20,1.5,a1,(4.5,0.05)] {};
	\aboxl[60,20,1.5,a2,(5.5,0.05)] {\textcolor{black}{$I+1$}};
	
	\node at (3,1.15) {$p_{c,in}^{n}$};
	\node at (4.8,1.15) {$p_{c,out}^{n}$};
	\node at (3,0.8) {$m_{c,in}^{n}$};
	\node at (4.8,0.8) {$m_{c,out}^{n}$};
	\end{tikzpicture}
	\caption{Compressor and control volumes}
	\label{fig:Compressor}
\end{figure}
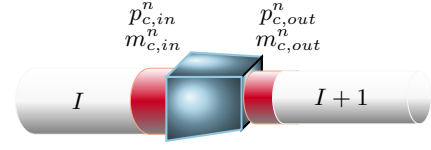

Instead of the pipelines, nodes can also be connected by compressors which is denoted by the set $\mathcal{C}$. Similar to the pipe segments, each compressor $c$ has associated mass and pressure variables as shown in \fref{fig:Compressor}.
The discretized version of \eqref{eq:CompressorRatioRelationship}
is 
\begin{equation}
\label{eq:CompressorPressureDiscrete}
\kappa_c=\frac{p_{c,out}^n}{p_{c,in}^n},  \quad \forall c \in \mathcal{C},
\end{equation} 
and it is assumed that the compression ratio $\kappa$ is fixed for the time horizon of the optimization problem. The discretized compressor consumption \eqref{eq:CompressorConsumption} is 
\begin{equation}
\label{eq:CompressorConsumptionDiscrete}
m_{c,con}^n=K(m^n_{c,out})\big(\kappa_{c}^\gamma -1 \big), \quad \forall c \in \mathcal{C}.
\end{equation}
Finally, the conservation of mass at each  compressor station requires
\begin{equation}
\label{eq:CompressorMassBalanceDiscrete}
m^n_{c,in}=m^n_{c,con}+m^n_{c,out},  \quad \forall c \in \mathcal{C}.
\end{equation}

\subsubsection{Junctions}
\begin{figure}[t]
	\centering
	\begin{tikzpicture}
	\tikzset{xzplane/.style={canvas is xz plane at y=#1, thick,shading=ball,ball color=white,draw=black}};
	\tikzset{yzplane/.style={canvas is yz plane at x=#1,thick,shading=ball, ball color=white,draw=black}};
	\tikzset{xyplane/.style={canvas is xy plane at z=#1,thick,shading=ball,ball color=white,draw=black}};
	
	\small
	
	\aboxl[60,25,2.5,a1,(0.2,0)] {\textcolor{black}{$I$}}; 
	\aboxm[30,25,2.5,a2,(1.2,0)] {};
	\begin{scope}[shift={(1.7 cm,-0.2 cm)}]
	\shade[ball color=mycolor5] (0.3,0.2) circle (4.0ex);
	\end{scope}
	\aboxmv[20,25,0.8,a2,(2,0.6)] {};
	\aboxrv[40,25,0.8,a2,(2,1.5)]{\textcolor{black}{$K$}};
	\aboxm[30,25,0.8,a2,(2.8,0)] {};
	\aboxr[60,25,0.8,a2,(4,0)] {\textcolor{black}{$J$}};
	\node at (1.2,-.8) {$x_{i+1}$};
	\node at (1.1,0.7) {$x_{k-1}$};
	\node at (3.,-.8)  {$x_{j-1}$};
	
	\end{tikzpicture}
	\caption{Typical network junction}
	\label{fig:Junction}
\end{figure}
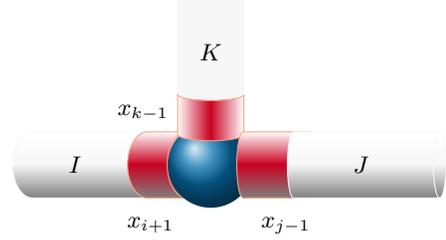

At the junction between two or more pipes or at the interface between discretized sections of a pipe (see \fref{fig:Junction}), the conservation of the mass flux has to be satisfied along with any injections $s_i^n$ or withdrawals $d_i^n$ from the network at that junction. This is achieved by

\begin{align}
	\label{eq:junctionmassbalance}
	s_i^n=d_i^n + \sum_{{v \in \mathcal{V}_i\cup\mathcal{W}_i}} m_v^n, && \forall i \in \mathcal{N}
\end{align}
where $\mathcal{V}_i$ is the set of all mass flows from pipes connected to node $i$ and $\mathcal{W}_i$ is the set of all flows of compressors connected to node $i$. Additionally, the nodal pressure value for each pipe connected to the junction should be equal, i.e. 
\begin{align}
	\label{eq:junctionbalance}
	p^n_{i+1}=p^n_{k-1}=p^n_{j-1}
\end{align}

\subsubsection{Initial Conditions}
In a real world setting, the state of the network at the starting time may be known, however if it is not available then a sensible initial condition for the network should be chosen. Throughout, we assume that the network starts from steady state which can be found by setting all the time derivatives to zero and solving the resulting nonlinear system of equations.

\subsection{Nonlinear system}

Following the fully implicit discretization of the governing equations (using
the usual finite volume method, with treatments as noted above), we express the nonlinear system of equations that describes the gas flow in a network as
\begin{align} 
	\label{eq:DiscretizedEquations} 
	\p^n(\x^n, \x^{n-1}, \u) =  \0
	\quad n ={1,2,\ldots,N} \end{align}
where $\p^n$ denotes the fully discretized, both in space and time, set of
partial differential equations as well as the junction conditions, compressor equations and initial conditions \eqref{eq:DiscreteMassBalance}-
\eqref{eq:junctionbalance}. Here $\x^n$ and $\u$ are
the network states (pressure and mass flow) at time step $n$ and controls (compressor ratios which are assumed to be constant over the optimization horizon), respectively. The
corresponding time step size is designated by $\Delta t^n$.
The Newton-Raphson method is used for the linearization of the nonlinear algebraic system, with the solution at the previous time step as the initial guess, similarly to \cite{Helgaker:2014},\cite{Abbaspour:2010}. 
The Newton iterations terminate when the maximum relative
norm of the residual is less than a specified tolerance  $\xi$.

\subsection{Newton-Raphson Continuation}
Discontinuous jumps of the compressor ratios cause large pressure jumps
which may prevent Newton from achieving convergence. In such cases, it is
necessary to adopt a Newton continuation approach whereby an initial solution to the nonlinear system is computed with all compressor ratios set to $\kappa = 1+\epsilon$, i.e., compression effects are marginal. 
Within Newton's iteration, the compressor ratios are gradually increased towards their true values, using the solution of the previous step as the initial guess.

\subsection{Compressor Flow Constraints}


Compressor flows should remain non-negative. This constraint can be introduced as an additional inequality constraint in the OGF problem. However, it is desirable to enforce the non-negativity of the compressor fluxes in the course of the simulation. For this purpose, in the event of flow reversal at one or more compressors, the compressors are turned off, i.e., the associated compressor ratios are set to one, and the Newton's iteration is restarted. The same process is repeated until convergence to a solution is achieved in which flow reversal does not occur.

\section{Optimal Gas Flow Problem}
\label{sec:OGF}
A gas network operator's main aim is to supply their customers with guaranteed output pressures at the customer nodes, while at the same time minimizing their costs.
The OGF problem therefore reads as
\begin{subequations}
	\label{eq:OGF}
	\begin{alignat}{2}
		& \underset{\u}{\text{minimize}} &&  J=\sum_{n=1}^N \Delta t^n \, \F^n \left ( \x^n, \u \right ) \label{eq:OGFobjective} \\
		& \text{subject to} \quad  
		&& \p^n(\x^n, \x^{n-1}, \u) = \0, \label{eq:OGFsimulation}\\
		&&& \u^{\text{min}} \leq \u \leq \u^{\text{max}}, \label{eq:OGFcompressor}\\
		&&& p^{\text{min}} \leq p_j^n \leq p^{\text{max}},  \quad\quad \forall j \in \mathcal{N}_0, \label{eq:OGFpressure}\\
		&&& \forall n = 1,2,\ldots,N. \nonumber
	\end{alignat}
\end{subequations}
The objective function \eqref{eq:OGFobjective} can be any nonlinear function of the controls and state variables but here we will use the total compression fuel use. The minimization of the compression cost is subject to the equations governing gas flow in networks \eqref{eq:OGFsimulation}.
In addition, the compressor ratios have to honor the bounds ($\kappa^\text{min},\kappa^\text{max}$). 
Similarly, \eqref{eq:OGFpressure} enforces bounds ($p^\text{min},p^\text{max}$) on pressure, which are given due to technical limits and contractual agreements. The pressure in the network decreases due to the friction inside the pipelines. It is therefore only necessary to apply this constraint at the original network nodes $\mathcal{N}_{0}$.



\subsection{Discrete adjoint formulation} \label{subsection:discreteAdjoint}

\begin{figure}[ht]
	\centering
	\scalebox{0.8}{\tikzstyle{Optimizer} = [rectangle, rounded corners, minimum width=8cm, minimum height=1cm,text centered, draw=black, fill=mycolor1]
\tikzstyle{Simulator} = [rectangle, minimum width=1cm, minimum height=1cm, text centered, text width=2cm, draw=black, fill=mycolor3]
\tikzstyle{Gradient} = [rectangle, minimum width=3cm, minimum height=1cm, text centered, text width=3cm, draw=black, fill=mycolor4]
\tikzstyle{arrow} = [thick,->,>=stealth]
\tikzstyle{Constraints} = [rectangle, minimum width=3cm, minimum height=1cm, text centered, text width=3cm, draw=black, fill=mycolor4]
\tikzstyle{arrow} = [thick,->,>=stealth]

\tikzstyle{Lagrange} = [rectangle, minimum width=2cm, minimum height=1cm, text centered, text width=2cm, draw=black, fill=mycolor2]
\tikzstyle{arrow} = [thick,->,>=stealth]

\tikzstyle{Adjoint} = [rectangle, minimum width=3cm, minimum height=1cm, text centered, text width=3cm, draw=black, fill=mycolor2]
\tikzstyle{arrow} = [thick,->,>=stealth]

\begin{tikzpicture}[node distance=2cm]

\node (tagOptimizer) [Optimizer]     at (4.5,2.5)     {Optimizer};
\node (tagSimulate)  [Simulator]     at (2., 0.5)   {Simulator};
\node (tagConstraint)[Constraints]   at (6,0.5) {Evaluate Objective \& Constraints};
\node (tagAdjoint)   [Adjoint]      at (6.75,-1.5) {Compute Adjoint gradient};
\node (tagLagrange)  [Lagrange]     at (2.25, -1.50)   {Calculate ${\blambda}^n$};

\draw [arrow] (2,2) -- (2,1); 
\draw [arrow] (tagSimulate) -- (tagConstraint);
\draw [arrow] (tagLagrange) -- (tagAdjoint);
\draw [arrow] (6,1) -- (6,2);
\draw [arrow] (2,0) -- (2,-1);
\draw [arrow] (8,-1) -- (8,2);

\node [rectangle,text width=5cm,anchor=west] at (1.4,1.5) {\small $\bf{u}^k$} ;
\node [rectangle,text width=5cm,anchor=west] at (3.2,0.7) {\small $\bf{u}^k,\bf{x}^k$} ;
\node [rectangle,text width=5cm,anchor=west] at (0.9,-0.5) {\small $\bf{u}^k,\bf{x}^k$} ;
\node [rectangle,text width=5cm,anchor=west] at (3.5,-1.2) {\small $\bf{u}^k,\bf{x}^k,{\blambda^n}$} ;
\node [rectangle,text width=5cm,anchor=west] at (5.4,1.5) { $\displaystyle J^k \enskip h(\bf{u}^k,\bf{x}^k)$} ;
\node [rectangle,text width=5cm,anchor=west] at (7.1,-0.5) { $\displaystyle \frac{\partial J^k}{\partial \bf{u}^k} \enskip \frac{\partial h^k}{\partial \bf{u}^k}$} ;
\end{tikzpicture}
}
	\caption{Objective $J$, constraints $h(\u,\x)$ and respective derivatives $\partial J/\partial \u, \partial h(\u,\x)/\partial \u$, are evalauted at each iteration $k$ of the optimizer}
	\label{fig:Optimization}
\end{figure}
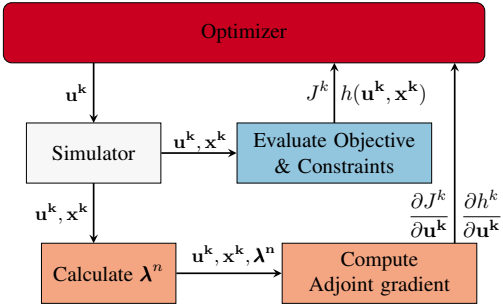

We adopt a discretize-then-optimize approach to solve the problem. 
The continuous problem is discretized and the states $\x$ are determined by solving the discretized gas flow equation~\eqref{eq:DiscretizedEquations} for compressor settings $\u$. The objective $J$, a function of $\x$ and $\u$, can then be evaluated. For minimizing $J$ with respect to $\u$, the gradient $\partial J/\partial u$ is needed and this is computed by the discrete adjoint formulation which will be described in this subsection. The advantage of this approach, in contrast to the 
optimize-then-discretize approach, is that it does not introduce errors in the gradient that grow with the time step size because it
does not use gradients for the continuous problem on the 
discrete implementation~\cite{Kourounis2014,Kourounis2015}. 
To describe the discrete adjoint formulation we start with the general form of a PDE constrained optimal control problem stated as
\begin{subequations}
	\label{eq:OGF_DAF}
	\begin{alignat}{2}
		& \underset{\u}{\text{minimize}} &&  J=\sum_{n=1}^N \Delta t^n \, \F^n \left ( \x^n, \u \right ) \label{eq:OGF_DAF1} \\
		& \text{subject to} \quad  
		&& \p^n(\x^n, \x^{n-1}, \u) = \0, \label{eq:OGF_DAF2}
	\end{alignat}
\end{subequations}
Now, we can introduce the augmented objective function $\myobj_A$ by `adjoining' 
the governing equations to the original objective function $\myobj$. The new objective $\myobj_A$ shares
the same extrema as $\myobj$, since $\p^n(\x^n,\x^{n-1},\u) = \0$, and is defined as
\begin{align}
	\label{eq:JAdefinition}
	\myobj_A = \fsum{n=1}{N}{\left ( \Delta t^n \F^n(\x^n, \u)
		+ (\blambda^{n})^{\transpose} \p^n (\x^n, \x^{n-1}, \u)  \right )},
\end{align}
where the vectors $\blambda^n$ are the Lagrange multipliers.
The maximum or minimum of $\myobj_A$ (and thus $\myobj$) is achieved when the first variation of $\myobj_A$ is 
zero ($\delta \myobj_A=0$). After performing some index shifting, and grouping
terms that are multiplied by the same variation ($\delta \x^n, \delta \x^n,
\delta \u$), $\delta \myobj_A$ can be written as
\begin{align}
	\label{eq:discreteAfterRegrouping}
	\delta \myobj_A &=
	\left (
	\Delta t^N \pder{\F^N}{\x^N}
	+ (\blambda^{N})^{\transpose}\pder{\p^N}{\x^N}
	\right ) \delta \x^N
	\nonumber \\
	&+\fsum{n=1}{N-1}{
		\left (
		\Delta t^n \pder{ \F^n}{\x^n}
		+(\blambda^{n+1})^{\transpose}\pder{\p^{n+1}}{\x^n}
		+(\blambda^{n})^{\transpose}\pder{\p^n}{\x^n}
		\right ) \, \delta \x^n
	} \nonumber \\
	&+\fsum{n=1}{N}{
		\left (
		\Delta t^n \pder{\F^n}{\u}
		+(\blambda^{n})^{\transpose}\pder{\p^n}{\u}
		\right ) \, \delta \u.
	}
\end{align}
In order to achieve $\delta \myobj_A=0$, we require $\delta \myobj_A / \delta \x_n=\0$ (for $n=1,2, \ldots, N$) and $\delta \myobj_A / \delta \u=\0$. To satisfy $\delta \myobj_A / \delta \x_n =\0$ for $n=1,2, \ldots, N$, we require that the Lagrange multipliers satisfy the following equations:
\begin{align}
	&\left (\pder{\p^n}{\x^n} \right )^\transpose  \blambda^n = -
	\left ( \pder{\p^{n+1}}{\x^n} \right )^\transpose \blambda^{n+1} - 
	\Delta t^n \left (\pder{\F^n}{\x^n} \right )^\transpose,
	\nonumber \\
	&\left( \pder{\p^N}{\x^N} \right)^\transpose \blambda^N = -
	\Delta t^N \left ( \pder{\F^N}{\x^N} \right )^\transpose.
	\label{eq:discreteODE}
\end{align}
The derivatives in \eqref{eq:discreteODE} can be evaluated using the solution of \eqref{eq:DiscretizedEquations}, 
and \eqref{eq:discreteODE} becomes a system of linear equations which can easily be solved.
With the resulting Lagrange multipliers the first and
second term of \eqref{eq:discreteAfterRegrouping} become zero and the gradient of the
objective function with respect to only the controls is

\begin{align}
	\label{eq:discreteGradient}
	\frac{\delta \myobj_A}{\delta \u} =
	\fsum{n=1}{N}{
		\left (
		\Delta t^n \pder{\F^n}{\u}
		+(\blambda^{n})^{\transpose}\pder{\p^n}{\u}
		\right ).
	}
\end{align}
Similarly, the gradient of any constraint $h(\u,\x)$ can be found by replacing $J$ with $h(\u,\x)$ in \eqref{eq:OGF_DAF1}.
Now that the gradients of the objective and constraints can be calculated with respect to only the compressor ratios 
we can adopt the reduced-space approach where the objective function is minimized
with respect to only control variables. The advantage of this method is that the 
the network constraints \eqref{eq:OGFsimulation} are not forwarded to the optimizer as
additional equality constraints because they are solved explicitly 
in order to find the gradients. A flow chart of the reduced space
approach is provided in \fref{fig:Optimization}. 

Summarizing, at each iteration of the optimization software a set of controls is produced, the simulation \eqref{eq:DiscretizedEquations} is solved and 
the results are used to compute the Lagrange multipliers \eqref{eq:discreteODE} for the objective and constraints (excluding the network constraints).
Lastly, the the Lagrange multipliers are used for the evaluation of the gradients using \eqref{eq:discreteGradient}. 
The objective function and the constraints are also evaluated and, together with their respective gradients, 
are supplied to the optimization software for it to complete its next iteration.

\subsection{Constraint Handling and Lumping}
\label{sec:constraint-nolumping}
Constraints that appear as simple bound constraints on the control variables such as \eqref{eq:OGFcompressor}, can be used directly as inputs to the optimizer. Constraints on state variables, such as the pressures, or functions of state variables, require the solution of the simulation for the evaluation of the  constraint and the solution of the adjoint problem for the evaluation of the gradient of the constraint with respect to the control variables, see~\cite{Kourounis2014},~\cite{Kourounis2015}.  

However, introducing pressure constraints for multiple nodes over multiple timesteps as individual inequality constraints may lead to an excessively large
number of constraints for realistic networks. This can have an adverse effect on the convergence of the optimizer and the overall run-time performance of the OGF problem. Therefore, we will use a constraint lumping approach to reduce the total number of constraints for the optimizer. The OGF problem for the original formulation, i.e., with no lumping, (OGF-NL) in this case reads
%
%
\[
\probNL{OGF-NL}{\u}{\displaystyle \myobj =
	\sum_{n=1}^N \Delta t^n \, \F^n \left ( \x^n, \u \right ) }
{\p^n(\x^n, \x^{n-1}, \u) = \0,}
{\u^{\text{min}} \leq \u \leq \u^{\text{max}},}
{p^{\text{min}} \le p^n_j \le p^{\text{max}},}
{\forall n=1,\ldots,N.}
{\forall j \in \mathcal{N}_0,}
\]
The number of constraints specified in the optimal
control problem are $2 |\mathcal{N}_0| \cdot N$, where $N$ is the number of time steps and $|\mathcal{N}_0|$
is the number of pressure nodes where the pressure should remain bounded.
Computing the gradient for each one of these constraints, requires the evaluation of the Lagrange multipliers corresponding to each constraint at every time step.  
Since the solution of \eqref{eq:discreteODE} requires the solution of a linear system at each timestep, in total $|\mathcal{N}_0| N (N+1)$ linear systems have to be solved, and this is in addition to each simulation that is needed to evaluate the Jacobian of the constraints. This may be computationally intractable for realistically sized problems. Nevertheless, this approach provides significantly higher flexibility to the optimizer in achieving feasible solutions, since the optimizer can manipulate pressures for every single node at the particular time steps where feasibility is violated.

A viable alternative is to introduce 
a single nonlinear constraint ~\cite{Kourounis2014,Kourounis2015}, satisfaction of which would 
guarantee that all pressure bounds are honored for all the pressure variables
of interest over all time steps. To illustrate the methodology let us assume
only upper bounds for the pressure variables
\begin{align}
	p^n_j \leq p_j^{\text{max}}, \quad \forall j \in \mathcal{N}_0, \quad n=1,\ldots,N,
\end{align}
where $p^n_j$ is the pressure 
defined at node $j$ at time step $n$ and $p_j^{\text{max}}$ is a specified maximum pressure at node
$j$.
All these constraints will be honored if the constraint
\begin{align}
	\max_{j,n}\{{p_j^n}/{ p_j^\text{max}}\} \leq 1, \quad \forall j \in \mathcal{N}_0, \quad n=1,\ldots,N,
\end{align}
is satisfied.
However, the $\max$ function is a non-differentiable function and thus, it
cannot be used to provide gradient information. Therefore, a differentiable approximation of the max function is introduced instead, specified as 
\begin{align}
	\label{eq:maxapprox}
	\max_{j,n}\bigg\{\frac{p^n_j}{p_j^\text{max}}\bigg\} \approx M \left (\frac{p^n_j}{p_j^\text{max}} \right ) = \alpha \log \left (\sum_{j=1}^{|\mathcal{N}_0|}\sum_{n=1}^{N} e^{\frac{p^n_j}{p_j^\text{max}}\frac{1}{\alpha}} \right )
\end{align}
with $\alpha=0.002$, a parameter introduced to prevent numerical overflow of the exponential terms. The approximation of the max becomes more accurate with decreasing values of $\alpha$. Lower values of $\alpha$ however, result in steep constraint gradients, which in turn may delay significantly the infeasibility reduction, since the initial guess suggested to the optimizer is usually  infeasible. The approximation of $\max$ in \eqref{eq:maxapprox} always
slightly overestimates the observed maximum, so if the approximation of max in  \eqref{eq:maxapprox} is bounded from above, the same is true for the constraint \eqref{eq:OGFpressure}.

The same process is followed for  obtaining a smooth approximation of the $\min$ function through the formula
\begin{align}
	&\min_{j,n} \{ p^n_j/p_{j}^\text{min} \} \approx \mu( p^n_j/p_{j}^\text{min}) =-M( p^n_j/p_{j}^\text{min}). 
\end{align}
The approximation of the $\max$ function by smooth functions is known in the literature as {\it constraint lumping} and it can be applied in different ways. In \eqref{eq:maxapprox}, lumping is applied both in time and space. We will refer to this approach as full lumping. The associated fully lumped OGF problem (OGF-FL) reads
%
%
%
\[
\probTL{OGF-FL}{\u}{\displaystyle \myobj =
	\sum_{n=1}^N \Delta t^n \, \F^n \left ( \x^n, \u \right ) }
{\p^n(\x^n, \x^{n-1}, \u) = \0, }
{\u^{\text{min}} \leq \u \leq \u^{\text{max}},}
{M(p_{j}^n/p_{j}^\text{max}) \le 1,}
{\mu(p_{j}^n/p_{j}^\text{min}) \geq 1 ,}
{\forall n=1,\ldots,N,}
{\forall j \in \mathcal{N}_0}
\]
where $M(\cdot)$ and $\mu(\cdot)$ are the approximation of the maximum and minimum values of all pressure values respectively.

Lumping can also be performed at each timestep for all nodes of the network, i.e., lump the pressure constraints in space. The space lumped OGF problem (OGF-SL) reads 
%
\[
\probTL{OGF-SL}{\u}{\displaystyle \myobj =
	\sum_{n=1}^N \Delta t^n \, \F^n \left ( \x^n, \u \right ) }
{\p^n(\x^n, \x^{n-1}, \u) = \0, }
{\u^{\text{min}} \leq \u \leq \u^{\text{max}},}
{M^n(\mathbf{p}_j/p_{j}^\text{max}) \le 1,}
{\mu^n(\mathbf{p}_j/p_{j}^\text{min}) \geq 1,}
{\forall n=1,\ldots,N,}
{\forall j \in \mathcal{N}_0}
\]
where $\mathcal{M}^n(\cdot)$ and $\mu^n(\cdot)$ are the approximation of the maximum and minimum values for all of the nodal pressure values at timestep $n$ respectively obtained through \eqref{eq:maxapprox}.
Finally, lumping can be performed in time, leading to the following definition of the time lumped (OGF-TL) problem
%
%
\[
\probTL{OGF-TL}{\u}{\displaystyle \myobj =
	\sum_{n=1}^N \Delta t^n \, \F^n \left ( \x^n, \u \right ) }
{\p^n(\x^n, \x^{n-1}, \u) = \0, }
{\u^{\text{min}} \leq \u \leq \u^{\text{max}},}
{M_j(\mathbf{p}^n/p_{j}^\text{max}) \le 1,}
{\mu_j(\mathbf{p}^n/p_{j}^\text{min}) \geq 1 ,}
{\forall n=1,\ldots,N,}
{\forall j \in \mathcal{N}_0}
\]
where $\mathcal{M}_{j}(\cdot)$ and $\mu_{j}(\cdot)$ are the approximation of the maximum and minimum values for all of the pressure values at node $j$ over all timesteps.
It is important to recognize that the approach used for constraint lumping can impact the convergence of the optimizer. Bound constraints on the controls \eqref{eq:OGFcompressor} do not require any special treatment as they are readily handled by the optimizer.

\section{Results}
\label{sec:Results}
We now present results for four different cases of increased complexity using the previously introduced formulations and approaches. 

\subsection{Benchmark Cases}
Details about the gas networks employed for our 
test cases are provided in Table~\ref{table:cases}. The aim is to
investigate the robustness of the proposed constraint-handling approaches as well as 
their runtime performance. Our investigation also considers various optimizers, more precisely IPOPT, a primal-dual interior-point method~\cite{IPOPT2005,IPOPT2006}, SNOPT, a sequential quadratic programming method~\cite{SNOPT}, and MATLAB\textregistered's interior point and SQP methods. Since each optimization method adopts a different approach for 
handling inequality constraints or for enforcing feasibility when the initial guess is infeasible, we expect this study, without claiming completeness, to reveal the most robust optimization methods for the OGF problem.

\begin{table} [!h]
	\centering
	\renewcommand{\arraystretch}{1.3}
	\caption{Benchmark cases}
	\label{table:cases}
	\begin{tabular}{cccccc}
		\toprule
		Case       & Nodes & Pipes & Compressors &Supply & Demand\\
		&       &       &             &Scaling& Scaling\\ \midrule
		GasLib-24  & 24    &  19   &  3          &  4.3         & 4.3\\
		GasLib-40  & 40    &  39   &  6          &  1.1         & 1.1\\
		GasLib-134 & 134   &  86   &  1	         & 1.2          & 0.72\\
		GasLib-135 & 135   & 141   & 29	         &  2.3         &  2.3\\
		\bottomrule
	\end{tabular}
\end{table}

The networks are taken from GasLib~\cite{GasLib}, which provides data for steady state analysis of gas networks. The gas loads $L(t)$ were chosen to be a sinusoidal function around the initial Load $L_0$ for the simulation period ($T$)
\begin{align}
	L(t)=L_{0}\left (1+0.2\sin{\left (\frac{t}{2\pi  T}\right )
	} \right).
\end{align}
In order to obtain actionable pressure drops in the network, all the initial loads provided by GasLib are scaled according to the values in Table \ref{table:cases}. The first source node of each network is set as a slack pressure node with a normalized value of 1 p.u. and the remaining source nodes are treated the same as the load nodes with scaling again according to Table \ref{table:cases}.

For all networks, the lower bound on pressure is set to $p^{\text{min}} = 0.7$ and the upper bound to $p^{\text{max}} = 1.1$ for all nodes in the network. The simulation period is set to $T=24$ Hours, with a constant time step size of $\delta t = 10$ minutes. The compressors are all modelled with the same values for the parameters of $K=0.1$ and $\gamma =1.2$. The lower and upper bounds on the compressor ratios are 1 and 1.2, respectively. The Newton-Raphson tolerance $\xi$ is set to $10^{-10}$.

\subsection{Discretization}
The spatial grid is obtained from the pipe network by subdividing each pipe into $N_h$ segments of equal size. The value of $N_h$ is chosen such that the discretization error is sufficiently small for both the objective function and the binding constraints. Keeping the values of the compressor ratios fixed, the objective function values and pressures are computed for all networks for different values of $N_h$ and the error is calculated with respect to a case with sufficiently large $N_h=20$ and time step size $\delta t=1$ minute. 

\begin{figure}[tb]
	\scalebox{0.9}{\begin{tikzpicture}

\begin{axis}[
width=0.82\columnwidth,
height=0.506\columnwidth,
at={(0.0in,0.0in)},
scale only axis,
separate axis lines,
every outer x axis line/.append style={,font=\footnotesize},
every x tick label/.append style={font=\footnotesize},
xmin=2,
xmax=14,
xtick pos=left,
ytick pos=left,
xlabel={Number of sections per pipe $N_h$},
every outer y axis line/.append style={,font=\footnotesize},
every y tick label/.append style={font=\footnotesize},
ymode=log,
ymin=0.0000001,
ymax=0.1,
max space between ticks=300pt,
try min ticks=3,
yminorticks=true,
max space between ticks=25,
xminorgrids=true,
xmajorgrids=true,
ymajorgrids=true,
ylabel={$\epsilon_{r}(N_{h})$},
axis background/.style={fill=none},
legend style={at={(0.5,1.2)},anchor=north,align=right,legend columns=7,legend cell align=left,draw=none,font=\footnotesize},
axis background/.style={
			shade,top color=mycolor3,bottom color=white},
xlabel/.append style={font=\scriptsize},
ylabel/.append style={font=\scriptsize},
]

\addplot[color=mycolor1,line width=1.5pt]
table[x=No_Sections,y=Obje_24_Rel]{Convergence.dat};
\addlegendentry{24};

\addplot[color=mycolor1,line width=1.5pt,dashed,forget plot]
table[x=No_Sections,y=Pe_24_Rel]{Convergence.dat};

\addplot[color=mycolor2,line width=1.5pt]
table[x=No_Sections,y=Obje_40_Rel]{Convergence.dat};
\addlegendentry{40};

\addplot[color=mycolor2,line width=1.5pt,dashed,forget plot]
table[x=No_Sections,y=Pe_40_Rel]{Convergence.dat};

\addplot[color=mycolor5,line width=1.5pt]
table[x=No_Sections,y=Obje_134_Rel]{Convergence.dat};
\addlegendentry{134};

\addplot[color=mycolor5,line width=1.5pt,dashed,forget plot]
table[x=No_Sections,y=Pe_134_Rel]{Convergence.dat};

\addplot[color=mycolor4,line width=1.5pt]
table[x=No_Sections,y=Obje_135_Rel]{Convergence.dat};
\addlegendentry{135};

\addplot[color=mycolor4,line width=1.5pt,dashed,forget plot]
table[x=No_Sections,y=Pe_135_Rel]{Convergence.dat};
\addlegendimage{}
\addlegendentry{$J_h$}
\addlegendimage{dashed}
\addlegendentry{min($p_h$)}

\end{axis}

\end{tikzpicture}%
}
	\caption{Absolute relative error $\epsilon_r$ of $J_h$ and $p_h$ for different values of $N_h$ for the considered systems (indicated by the number of nodes)}
	\label{fig:ConvergencePlotRelative}
\end{figure}
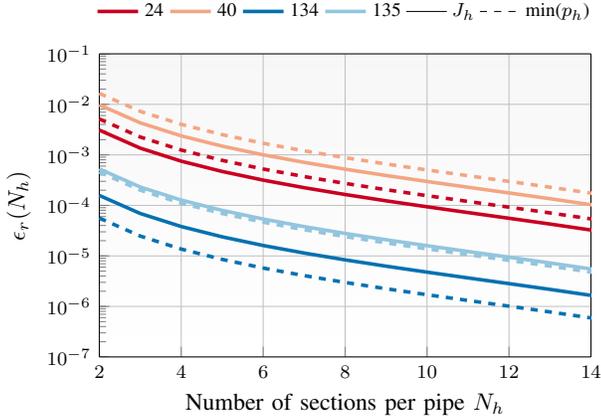

In \fref{fig:ConvergencePlotRelative}, the error in the objective varies with the network but it decreases with increasing $N_h$ as expected. The same is true for the error in the network minimum pressure although it is already quite small even for two subdivisions of each pipe, i.e., $N_h=2$. Choosing $N_h=10$ sections per pipe ensures that the error in the objective function, due to discretization, is kept below 0.1\% for all networks.

\subsection{Optimization Results}
Table \ref{table:Performance} 
summarizes the number of iterations required by each of the optimizers until convergence and the running time of the optimizer for all different constraint lumping methods. If convergence was achieved, all the optimizers converged to the same solution. Table~\ref{table:Objective} lists the objective values and minimum pressures for each of the constraint lumping methods. In all cases, the constraint lumping significantly reduces the total time for the optimization while there is usually not a significant change to the number of required iterations. The reductions in computation time become more apparent with larger networks and so does the performance of the optimizers. The FMINCON optimizers provided in MATLAB\textregistered~have the best performance in terms of iterations and running time for the first three examples. However, for the largest network, the FMINCON-SQP method fails in all cases, while the FMINCON-IP method can only solve the OGF-NL and OGF-FL problems at a very high computational cost, demonstrating that these methods are not applicable for large scale problems. 

Constraint lumping provides an approximation of the true constraints of the network, however, as with all approximations, it introduces an error. But this error in the pressure is never more than 1\% and the calculated pressure is always greater than the actual minimum pressure, i.e. the approximation is conservative. Due to the over-approximation of the pressures in the network, more compression is performed in the network resulting in a higher objective function value in the cases with constraint lumping. Due to the nonlinear effects in the network, the error in the objective can be up to 24\% for the full lumping case. In general, the space lumping performs the best in terms of computation time and error providing at least a 6-fold speedup in runtime performance over the OGF-NL method. Since the performance of OGF-NL increases with an increasing number of timesteps, and pressure nodes, we expect that the OGF-SL problem will provide higher speedups for larger number of timesteps or larger networks.
\begin{table} [!h]
	\centering
	\renewcommand{\arraystretch}{1.3}
	\caption{Optimizers iterations (time in minutes)}
	\label{table:Performance}
	\begin{tabular}{ccccc} 
		\toprule
		Optimizer   & OGF-NL    & OGF-TL   & OGF-SL   & OGF-FL\\ 
		\cline{2-5}
		& \multicolumn{4}{c}{ GasLib-24 }\\ 
		\cline{2-5}
		IPOPT &12(8.80) &8(0.50) &15(1.46) &9(0.36)\\
		SNOPT &10(10.89) &11(1.14) &10(1.54) &12(0.87)\\
		FMINCON-IP &9(6.40) &6(0.37) &7(0.70) &6(0.25)\\
		FMINCON-SQP &5(3.79) &5(0.32) &5(0.53) &5(0.21)\\
		\cline{2-5} 
		& \multicolumn{4}{c}{ GasLib-40 }\\ 
		\cline{2-5}
		IPOPT &15(33.17) &10(1.40) &13(2.45) &9(0.66)\\
		SNOPT &8(29.90) &11(2.59) &8(2.51) &11(1.44)\\
		FMINCON-IP &13(27.47) &9(1.19) &11(1.96) &7(0.51)\\
		FMINCON-SQP &3(7.86) &3(0.48) &3(0.66) &3(0.26)\\
		\cline{2-5} 
		& \multicolumn{4}{c}{ GasLib-134 }\\ 
		\cline{2-5}
		IPOPT &10(390.46) &7(8.95) &8(5.79) &6(0.90)\\
		SNOPT &6(357.89) &8(15.31) &6(6.63) &10(2.36)\\
		FMINCON-IP &7(254.45) &5(6.01) &5(3.54) &5(0.73)\\
		FMINCON-SQP &3(130.84) &3(4.03) &3(2.34) &3(0.49)\\
		\cline{2-5} 
		& \multicolumn{4}{c}{ GasLib-135 }\\ 
		\cline{2-5}
		IPOPT &43(2953.75) &37(81.57) &38(52.85) &47(23.44)\\
		SNOPT &17(2050.20) &17(59.80) &17(35.96) &17(11.86)\\
		FMINCON-IP &65(10449.33) &-- &69(132.01) &--\\
		FMINCON-SQP &26(7522.22) &25(221.35) &26(131.29) &25(44.83)\\
		\bottomrule 
	\end{tabular}
	
\end{table}
\begin{table} [!h]
	\centering
	\renewcommand{\arraystretch}{1.3}
	\caption{Optimization objective and pressure}
	\label{table:Objective}
	\begin{tabular}{ccccc} 
		\toprule
		& OGF-NL & OGF-TL   & OGF-SL   & OGF-FL\\ 
		\cline{2-5}
		& \multicolumn{4}{c}{ GasLib-24 }\\ 
		\cline{2-5}
		Objective &1766 &1832 &1766 &1832\\
		Pressure &0.7000 &0.7047 &0.7000 &0.7047\\
		\cline{2-5} 
		& \multicolumn{4}{c}{ GasLib-40 }\\ 
		\cline{2-5}
		Objective &352 &368 &352 &369\\
		Pressure &0.7000 &0.7041 &0.7001 &0.7042\\
		\cline{2-5} 
		& \multicolumn{4}{c}{ GasLib-134 }\\ 
		\cline{2-5}
		Objective &250 &283 &280 &311\\
		Pressure &0.7000 &0.7052 &0.7047 &0.7096\\
		\cline{2-5} 
		& \multicolumn{4}{c}{ GasLib-135 }\\ 
		\cline{2-5}
		Objective &913 &1069 &937 &1095\\
		Pressure &0.7000 &0.7056 &0.7009 &0.7065\\
		\bottomrule 
	\end{tabular}	
\end{table}

\begin{figure} [ht]
	\centering
	\begin{tabular}{c}
		-0.19\% \hfill{~} 0.14\%
		\\
		\includegraphics[width=0.8\columnwidth,angle=0]{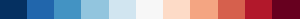}  
		\\   
		\includegraphics[height=0.5\columnwidth,width=0.8\columnwidth,angle=0]{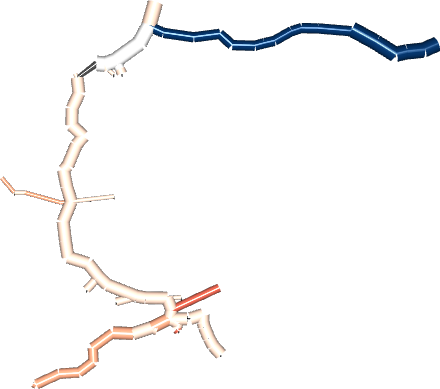}
	\end{tabular}
	\caption{GasLib-134, edge mass flux percent error between the solutions of 
		the OGF-NL and OGF-TL problems}
	\label{fig:Example2}
\end{figure}
\begin{figure} [ht]
	\centering
	\begin{tabular}{c}
		-0.84\% \hfill{~} 0.85\%
		\\
		\includegraphics[width=0.8\columnwidth,angle=0]{ColorBarRgButhin.png}
		\\
		\includegraphics[height=0.6\columnwidth,width=0.8\columnwidth,angle=0]{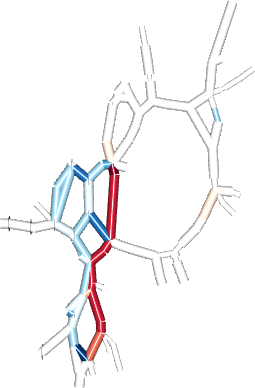}
	\end{tabular}
	\caption{GasLib-135, edge mass flux percent error between the solutions of 
		the OGF-NL and OGF-SL problems}
	\label{fig:Example3}
\end{figure}

Finally, in \fref{fig:Example2}, we depict the percentage error in the mass flux from the optimal solution of the OGF-NL problem and the OGF-TL problem, computed as $m^*_{NL} - m^*_{TL}$, for the GasLib-134 benchmark. Similarly, \fref{fig:Example3} depicts the percentage error in the nodal pressure from the optimal solution of the OGF-NL problem and OGF-SL problem, computed as the $p^*_{NL} - p^*_{SL}$, for the GasLib-135 benchmark. In both cases, the percentage error depends on the location in the network, however it is in general negligibly small.

\section{Conclusions}
\label{sec:conclusion}
An efficient adjoint gradient-based optimization framework was 
presented for the minimization of compression cost subject to the
transient isothermal gas flow equations and inequality constraints on
the pressure. 
The gradients from the discrete adjoint formulation allow for a rapid convergence, in less than 20 iterations on average, for all optimizers and for all benchmark cases studied. 
Our numerical analysis demonstrated that lumping-based constraint-handling methods
can accelerate the solution process of the OGF problem without deteriorating significantly
the optimality of the solution. Full lumping tends to
introduce marginally suboptimal solutions since the maximum or minimum
of the pressure are usually slightly overestimated or underestimated, respectively.
Lumping over space or over time provide tighter approximations of
both the maximum and minimum leading to smaller errors in the objective
function compared with the case where no lumping is performed. For the
largest benchmark case GasLib-135, OGF-SL allowed for a 50-fold increase in computation time compared to the case where no
lumping is performed. Higher 
speedups are expected for larger networks, and for higher number of timesteps.
Overall, the numerical investigation revealed that the adjoint gradient-based
optimization framework with the proposed constraint-lumping methods,
leads to feasible solutions for the continuous problem and it is also
practical from an operational standpoint. 



%



\ifCLASSOPTIONcaptionsoff
  \newpage
\fi

\bibliographystyle{IEEEtran}






\end{document}